\documentclass[10pt,conference]{IEEEtran}
\usepackage{cite}
\usepackage{amsmath,amssymb,latexsym,mathrsfs,wasysym,marvosym}
\usepackage{amssymb}
\usepackage[pdftex]{graphicx,color}
\usepackage{psfig}
\usepackage{psfrag}
\usepackage{subfigure}
\usepackage{url}
\usepackage{array,multirow}
\usepackage{fontenc,times}
\usepackage{pxfonts}
\usepackage{nohyperref} 
\makeatletter

\DeclareMathAlphabet{\eurm}{U}{eur}{m}{n}
\DeclareMathAlphabet{\mathbsf}{OT1}{cmss}{bx}{n}
\DeclareMathAlphabet{\mathssf}{OT1}{cmss}{m}{sl}
\DeclareMathAlphabet{\mathcsf}{OT1}{cmss}{sbc}{n}

\DeclareSymbolFont{bsfletters}{OT1}{cmss}{bx}{n}  
\DeclareSymbolFont{ssfletters}{OT1}{cmss}{m}{n}
\DeclareMathSymbol{\bsfGamma}{0}{bsfletters}{'000}
\DeclareMathSymbol{\ssfGamma}{0}{ssfletters}{'000}
\DeclareMathSymbol{\bsfDelta}{0}{bsfletters}{'001}
\DeclareMathSymbol{\ssfDelta}{0}{ssfletters}{'001}
\DeclareMathSymbol{\bsfTheta}{0}{bsfletters}{'002}
\DeclareMathSymbol{\ssfTheta}{0}{ssfletters}{'002}
\DeclareMathSymbol{\bsfLambda}{0}{bsfletters}{'003}
\DeclareMathSymbol{\ssfLambda}{0}{ssfletters}{'003}
\DeclareMathSymbol{\bsfXi}{0}{bsfletters}{'004}
\DeclareMathSymbol{\ssfXi}{0}{ssfletters}{'004}
\DeclareMathSymbol{\bsfPi}{0}{bsfletters}{'005}
\DeclareMathSymbol{\ssfPi}{0}{ssfletters}{'005}
\DeclareMathSymbol{\bsfSigma}{0}{bsfletters}{'006}
\DeclareMathSymbol{\ssfSigma}{0}{ssfletters}{'006}
\DeclareMathSymbol{\bsfUpsilon}{0}{bsfletters}{'007}
\DeclareMathSymbol{\ssfUpsilon}{0}{ssfletters}{'007}
\DeclareMathSymbol{\bsfPhi}{0}{bsfletters}{'010}
\DeclareMathSymbol{\ssfPhi}{0}{ssfletters}{'010}
\DeclareMathSymbol{\bsfPsi}{0}{bsfletters}{'011}
\DeclareMathSymbol{\ssfPsi}{0}{ssfletters}{'011}
\DeclareMathSymbol{\bsfOmega}{0}{bsfletters}{'012}
\DeclareMathSymbol{\ssfOmega}{0}{ssfletters}{'012}

\newtheorem{theorem}{\textbf{Theorem}}

\newtheorem{definition}{\textbf{Definition}}

\newtheorem{remark}{Remark}
\newcommand{\dv}{\mathbf} 
\newcommand{\mc}{\mathcal} 

\newcommand{\qed}{\nobreak \ifvmode \relax \else
      \ifdim\lastskip<1.5em \hskip-\lastskip
      \hskip1.5em plus0em minus0.5em \fi \nobreak
      \vrule height0.5em width0.5em depth0.25em\fi}

\begin{document}
\fontencoding{OT1}\fontsize{9.4}{11.25pt}\selectfont
\title{Achievable Regions for Interference Channels with Generalized and Intermittent Feedback\\}

\author{Abdellatif Zaidi$\:^{\dagger}$ \vspace{0.3cm} \thanks{This work has been supported by the European Commission in the framework of the FP7 Network of Excellence in Wireless Communications (NEWCOM\#).}\\
$^{\dagger}$ Universit\'e Paris-Est Marne La Vall\'ee, Champs-sur-Marne 77454, France\\
abdellatif.zaidi@univ-mlv.fr
}

\maketitle

\begin{abstract}
In this paper, we first study a two-user interference channel with generalized feedback. We establish an inner bound on its capacity region. The coding scheme that we employ for the inner bound is based on an appropriate combination of Han-Kobayash rate splitting and compress-and-forward at the senders. Each sender compresses the channel output that is observes using a compression scheme that is \`a-la Lim \textit{et al.} noisy network coding and Avestimeher \textit{et al.} quantize-map-and-forward. Next, we study an injective deterministic model in which the senders obtain output feedback only intermittently. Specializing the coding scheme of the model with generalized feedback to this scenario, we obtain useful insights onto effective ways of combining noisy network coding with interference alignment techniques. We also apply our results to linear deterministic interference channels with intermittent feedback. 
\end{abstract}

\section{Introduction}\label{secI}

The interference channel (IC) models situations in which separate senders communicate with distinct destinations over a common channel. In this model, the signal transmitted by one sender constitutes interference in the eyes of other sender-receiver pairs. The study of the IC was initiated by Shannon \cite{C61} and further studied by Ahlswede \cite{A74} and Carleial \cite{C78}. The capacity region of the IC is still unknown, and the best achievable rate region to date is due to Han and Kobayashi \cite{HK81}.

It is well known that feedback does not increase the capacity of memoryless point-to-point communication channels. However, feedback can enlarge the capacity in multiuser channels by enabling statistical cooperation among the transmitters (see, e.g., \cite{D80} and references therein). For interference networks with feedback, the choice of an appropriate form of users cooperation for managing interference depends highly on the quality of feedback signals that the transmitters get. For example, when the feedback is perfect or noiseless users can perform partial decode-and-forward or variants of it; and this offers substantial rate gains. In this case, capacity gains can even be unbounded if the feedback is free of cost \cite{ST11}.  Partial decode-and-forward type cooperation can also be exploited for managing interference in certain noisy feedback settings \cite{YT11}. For other settings, impairments on the feedback links may render the feedback signals so weak that requiring each transmitter to decode all or part of other users messages may only lead to stringent rate constraints. In such settings, schemes in which senders perform compressions of the observed feedback signals may be more appropriate.

\begin{figure}[htpb]
\centering
\includegraphics[width=\linewidth]{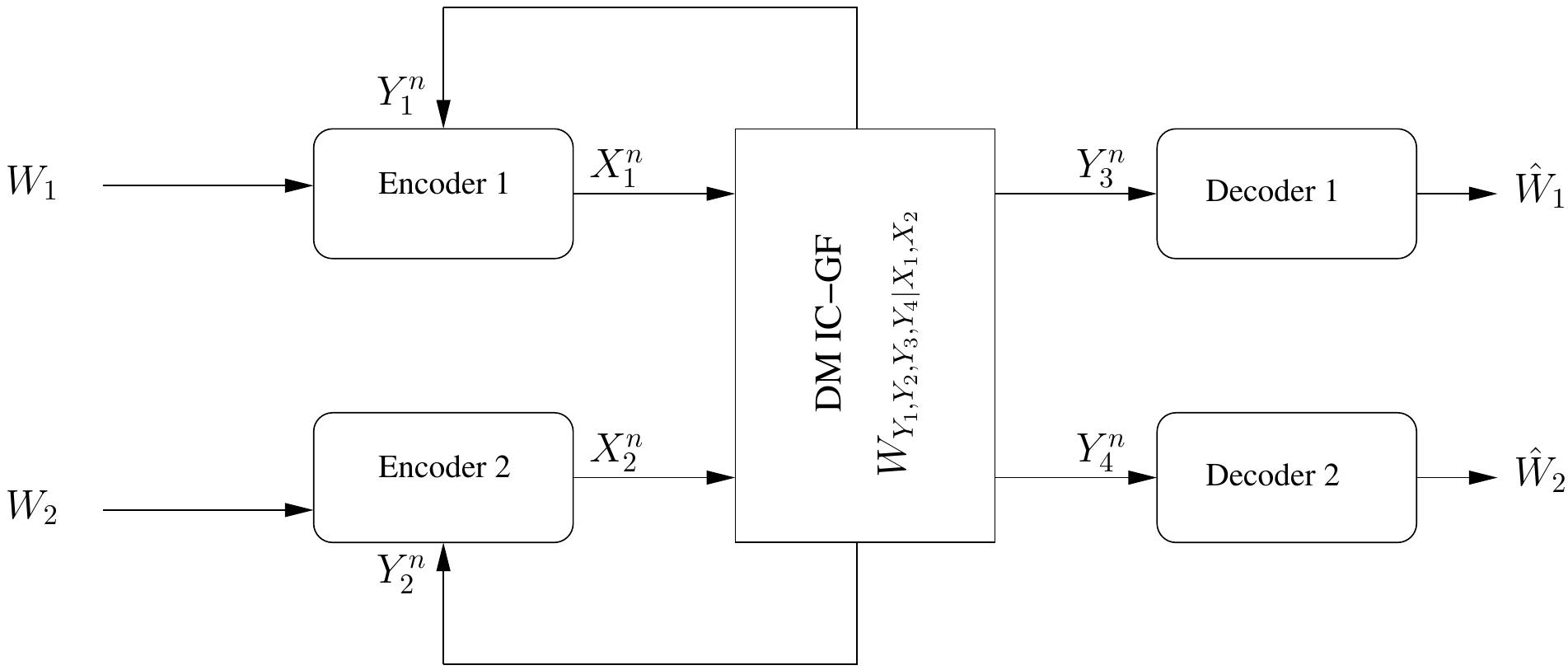}
\caption{Memoryless interference channel with generalized feedback (IC-GF) with two source-destination pairs.}
\label{fig-ic-gf}
\vspace{-0.5cm}
\end{figure}

In this paper, we first study a two-user memoryless interference channel with \textit{generalized} feedback (IC-GF). We establish an inner bound on its capacity region. The coding scheme that we employ for the proof of the inner bound is based on an appropriate combination of Block-Markov coding, Han-Kobayashi rate-splitting \cite{HK81} and compression at the encoders \cite{CG79}. An important ingredient of the coding scheme is how the compression is performed at the encoders. Note that since the destinations generally observe distinct outputs signals, for this model compressions that perform binning at the encoders and unique decoding of the compression indices, i.e., \`a-la Wyner-Ziv \cite{WZ76}, may result in rates that are limited by the worst side information. For this reason, two important features of the compression in our coding scheme are 1) standard compression without Wyner-Ziv binning and 2) non-explicit decoding of the compression indices. That is, the compression is \`a-la Lim \textit{et al.} noisy network coding \cite{H-LKGC11} or Avestimeher \textit{et al.} quantize-map-and-forward \cite{ADT11}. 

Next, we apply the results to El Gamal and Costa injective deterministic IC \cite{E-GC82}, with additional feedback given \textit{only intermittently} to the transmitters. In this model, at each time instant, each encoder either observes the previous channel output of its corresponding receiver or an erasure symbol, depending on a binary valued state variable. Aspects of intermittence have been studied in some related works in the literature, such as for a linear deterministic interference channel in \cite{KWD13,WSDV13} and for a multiaccess channel in \cite{KL13}. By specializing the aforementioned inner bound to this model, we show that the scheme reduces to one in which noisy-network coding is combined appropriated with interference alignment. 



\section{Problem Setup}\label{secII}

Consider the two sender-receiver (2-user) pair discrete memoryless interference channel with generalized feedback IC-GF $(\mc X_1{\times}\mc X_2, p(y_1,y_2,y_3,y_4|x_1,x_2),\mc Y_1{\times}\mc Y_2{\times}\mc Y_3{\times}\mc Y_4)$ shown in Figure~\ref{fig-ic-gf}. It consists of two input alphabets $\mc X_1$, $\mc X_2$, four output alphabets $\mc Y_1$, $\mc Y_2$, $\mc Y_3$ and $\mc Y_4$, and a collection of conditional probability mass functions $p(y_1,y_2,y_3,y_4|x_1,x_2)$ on $\mc Y_1{\times}\mc Y_2{\times}\mc Y_3{\times}\mc Y_4$. All input and output alphabets are assumed to be finite. Source $k$, $k=1,2$, has a message $W_k$ that it wants to transmit to Destination $k$. The messages $W_1$ and $W_2$ are independent random variables drawn uniformly from the sets $\mc W_1=\{1,\cdots,M_1\}$ and  $\mc W_2=\{1,\cdots,M_2\}$, respectively. The channel is modeled as a memoryless conditional probability distribution $W_{Y_1,Y_2,Y_3,Y_4|X_1,X_2}$ so that the law governing $n$-sequences of output letters is
\begin{align}
&P_{Y^n_1,Y^n_2,Y^n_3,Y^n_4|X^n_1,X^n_2}(y^n_1,y^n_2,y^n_3,y^n_4| x^n_1, x^n_2) \nonumber\\
&\quad = \prod_{i=1}^n W_{Y_1,Y_2,Y_3,Y_4|X_1,X_2}(y_{1,i},y_{2,i},y_{3,i},y_{4,i}|x_{1i},x_{2i}).
\end{align}

Receiver $k$, $k=1,2$, guesses the message $W_k$ that is intended to it using its channel output $Y^n_{k+2}$.

\begin{definition}\label{basic-definitions-strictl}
For positive integers $n$, $M_1$ and $M_2$, an $(M_1,M_2,n,\epsilon)$ code for the interference channel with generalized feedback consists of sequences of encoder mappings
\begin{align}
\phi_{k,i}: \mc W_k{\times}\mc Y^{i-1}_k \longrightarrow \mc X_k, \quad k=1,2, \:\: i=1,\hdots,n
\label{encoding-functions}
\end{align}
and decoding maps
\begin{align}
\psi_k : \mc Y^n_{k+2} \longrightarrow \mc \mc W_k, \quad k=1,2
\label{decoding-functions}
\end{align}
such that the maximum probability of error among the two decoders does not exceed $\epsilon$,
\begin{equation}
\max_{k\in\{1,2\}} \text{Pr}[\hat{W}_k \neq W_k] \le \epsilon 
\label{definition-probability-of-error}
\end{equation}
The rate of messages $W_1$ and $W_2$  are defined as
\begin{equation*}
R_1 = \frac{1}{n}\log M_1 \qquad \text{and} \qquad R_2 = \frac{1}{n}\log M_2,
\end{equation*}
respectively.
A rate pair $(R_1,R_2)$ is said to be achievable if for every $\epsilon > 0$ there exists an $(2^{nR_1},2^{nR_2},n,\epsilon)$ code for the channel $W_{Y_1,Y_2,Y_3,Y_4|X_1,X_2}$.  The capacity region $\mc C_{\text{ic-gf}}$ of the interference channel with generalized feedback (IC-GF) is defined as the closure of the set of achievable rate pairs.
\end{definition}

\vspace{0.1cm}

The results of this paper are only outlined. Detailed proofs can be found in \cite{Z14a,Z14b}.

\section{Interference Channel with Generalized Feedback}\label{secIII}

\subsection{Inner Bound}\label{secIII_subsecA}

Let $\mc P^{\text{in}}_{\text{ic-gf}}$ stand for the collection of all random variables $(Q,U_1,V_1,U_2,V_2,X_1,X_2,Y_1,Y_2,Y_3,Y_4)$ such that $Q$, $U_1$, $V_1$, $U_2$, $V_2$, $X_1$ and $X_2$ take values in finite alphabets $\mc Q$, $\mc U_1$, $\mc V_1$, $\mc U_2$, $\mc V_2$, $\mc X_1$ and $\mc X_2$, respectively, and satisfy
\begin{align}
&P_{Q,U_1,V_1,U_2,V_2,X_1,X_2,Y_1,Y_2,Y_3,Y_4}(q,u_1,v_1,u_2,v_2,x_1,x_2,y_1,y_2,y_3,y_4) \nonumber\\
&= P_Q(q)P_{U_1,X_1|Q}(u_1,x_1|q)P_{U_2,X_2|Q}(u_2,x_2|q)P_{V_1|U_1,Y_1,Q}(v_1|u_1,y_1,q)\nonumber\\
&\:\: {\cdot} P_{V_2|U_2,Y_2,Q}(v_2|u_2,y_2,q)W_{Y_1,Y_2,Y_3,Y_4|X_1,X_2}(y_1,y_2,y_3,y_4|x_1,x_2).
\label{measure-inner-bound-ic-gf}
\end{align}

The relations in \eqref{measure-inner-bound-ic-gf} imply that $V_1 \leftrightarrow U_1 \leftrightarrow X_1$, $V_2 \leftrightarrow U_2 \leftrightarrow X_2$ and $(U_1,U_2,V_1,V_2)  \leftrightarrow (X_1,X_2) \leftrightarrow (Y_1,Y_2,Y_3,Y_4)$ are  Markov chains.

Define $\mc R^{\text{in}}_{\text{ic-gf}}$ to be the set of all rate pairs $(R_1,R_2)$ such that there exist non-negative numbers $R_{10}$, $R_{20}$ satisfying
\begin{subequations}
\begin{align}
\label{inner-bound-ic-gf-constraint1}
&R_1 \leq I(U_1,V_1,X_1;Y_3,U_2,V_2|Q)-I(V_1;Y_1|U_1,Q)\\
&R_1+R_{20} \leq I(U_1,V_1,U_2,V_2,X_1;Y_3|Q) + I(U_2,V_2;U_1,V_1,X_1|Q)\nonumber\\
&\qquad \qquad - I(V_1;Y_1|U_1,Q) - I(V_2;Y_2|U_2,Q)\\
\label{inner-bound-ic-gf-constraint2}
&R_1-R_{10}+R_{20} \leq \min\big\{I(U_1,V_1,U_2,V_2,X_1;Y_3|Q)\nonumber\\
&\qquad \qquad - I(V_1;Y_1|U_1,Q),\:\:  I(U_2,V_2,X_1;Y_3|U_1,V_1,Q)\big\}\nonumber\\ 
&\qquad \qquad + I(U_2,V_2;U_1,V_1,X_1|Q) - I(V_2;Y_2|U_2,Q)\\
\label{inner-bound-ic-gf-constraint3}
&R_1-R_{10}  \leq \min\big\{I(U_1,V_1,X_1;Y_3|U_2,V_2,Q),\nonumber\\
&\qquad \qquad I(U_2,V_2,X_1;Y_3|U_1,V_1,Q)\big\} + I(U_2,V_2;U_1,V_1,X_1|Q) \nonumber\\
&\qquad \qquad - I(V_1;Y_1|U_1,Q) - I(V_2;Y_2|U_2,Q)\\
\label{inner-bound-ic-gf-constraint4}
&R_1-R_{10} \leq I(X_1;Y_3|U_1,V_1,U_2,V_2,Q) + I(U_2,V_2;U_1,V_1,X_1|Q)\\
\label{inner-bound-ic-gf-constraint5}
&R_2 \leq I(U_2,V_2,X_2;Y_4,U_1,V_1|Q)-I(V_2;Y_2|U_2,Q)\\
\label{inner-bound-ic-gf-constraint6}
&R_2+R_{10} \leq I(U_1,V_1,U_2,V_2,X_2;Y_4|Q) + I(U_1,V_1;U_2,V_2,X_2|Q)\nonumber\\
&\qquad \qquad - I(V_1;Y_1|U_1,Q) - I(V_2;Y_2|U_2,Q)\\
\label{inner-bound-ic-gf-constraint7}
&R_2-R_{20}+R_{10} \leq \min\big\{I(U_1,V_1,U_2,V_2,X_2;Y_4|Q)\nonumber\\
&\qquad \qquad - I(V_2;Y_2|U_2,Q),\:\: I(U_1,V_1,X_2;Y_4|U_2,V_2,Q)\big\}\nonumber\\
&\qquad \qquad + I(U_1,V_1;U_2,V_2,X_2|Q) - I(V_1;Y_1|U_1,Q)\\ 
\label{inner-bound-ic-gf-constraint8}
&R_2-R_{20}  \leq \min\big\{I(U_2,V_2,X_2;Y_4|U_1,V_1,Q), \nonumber\\
&\qquad \qquad I(U_1,V_1,X_2;Y_4|U_2,V_2,Q)\big\} + I(U_1,V_1;U_2,V_2,X_2|Q)\nonumber\\
&\qquad \qquad - I(V_1;Y_1|U_1,Q) - I(V_2;Y_2|U_2,Q)\\
&R_2-R_{20} \leq I(X_2;Y_4|U_1,V_1,U_2,V_2,Q) + I(U_1,V_1;U_2,V_2,X_2|Q)
\label{inner-bound-ic-gf-constraint9}
\end{align}
\label{inner-bound-ic-gf}
\end{subequations}
for some  joint distribution of the form \eqref{measure-inner-bound-ic-gf}.

\vspace{0.3cm}

\begin{theorem}\label{theorem-inner-bound-ic-gf}
The capacity region of the interference channel with generalized feedback satisfies
\begin{equation}
\mc R^{\text{in}}_{\text{ic-gf}} \subseteq \mc C_{\text{ic-gf}}.
\end{equation}
\end{theorem}

\vspace{0.3cm}

\textbf{Proof:} An outline proof of the coding scheme that we use for the proof of Theorem~\ref{theorem-inner-bound-ic-gf} is given in Section~\ref{secV}. The associated error analysis may be found in \cite{Z14b}.

\subsection{Comments}\label{secIII_subsecA}

\begin{remark}\label{remark-main-idea-inner-bound-ic-gf}
The proof of Theorem~\ref{theorem-inner-bound-ic-gf} is based on a Block-Markov coding scheme in which each encoder sends a compressed version of its output observation to the receivers, in addition to public and private messages obtained through rate splitting \`a-la Han-Kobayashi \cite{HK81}. A key ingredient of our coding scheme is how the compression is performed at the encoders. Note that since the destinations generally observe distinct outputs signals, for this model compressions that perform binning at the encoders and unique decoding of the compression indices, i.e., \`a-la Wyner-Ziv \cite{WZ76}, may result in rates that are limited by the worst side information. For this reason, two important features of the compression in our coding scheme are 1) standard compression without Wyner-Ziv binning and 2) non-explicit decoding of the compression indices. That is, the compression is \`a-la Lim \textit{et al.} noisy network coding \cite{H-LKGC11} or Avestimeher \textit{et al.} quantize-map-and-forward \cite{ADT11}. More precisely, unlike the original compress-and-forward scheme by Cover and El Gamal \cite{CG79} where every information message is divided into blocks and different submessages are sent over these blocks and then decoded one at a time using the same codebook, here the \textit{entire} public and private messages are transmitted over \textit{all} blocks using codebooks that are generated independently, one for each block, and the decoding is performed simultaneously using all blocks. Also,  like \cite{H-LKGC11} and \cite{ADT11}, at each block the compression index of the output that is observed at the previous block is sent using standard rate distortion, not Wyner-Ziv binning. At the end of the transmission, Receiver $k$, $k=1,2$, uses its output from all blocks to perform simultaneous decoding of the public message $w_{k0}$ and private message $w_{kk}$, without uniquely decoding neither the public message sent by the other encoder nor the compression indices. 
\end{remark}

\begin{remark}\label{remark-auxiliary-random-variables-inner-bound-ic-gf}
In \eqref{inner-bound-ic-gf}, the random variable $Q$ serves as a time sharing random variable; the random variable $U_k$, $k=1,2$, carries the public message $W_{k0}$ sent by Encoder $k$ and the random variable $V_k$ carries a compression of the output $Y_k$. The transmission takes place in $B$ blocks. The message $W_k$, $k=1,2$, is divided into independent parts, a public message $W_{k0}$, sent at rate $R_{k0}$, and a private message $W_{kk}$, sent at rate $R_{kk}$.  At the beginning of block $i$, $i=1,\hdots,B$, Encoder $k$, $k=1,2$, does not compress only the output $\dv y_k[i-1]$ that is has observed in block $i-1$, but also the part of its input of the last block that carries the public message. More precisely, denote by $\dv u_k[i-1]:= \dv u_{k,i-1}(w_{k0},t_{k,i-2})$ the part of the input of Encoder $k$ that is sent in block $i-1$ and carries the public message $w_{k0}$ as well as the compression index $t_{k,i-2}$ of the output $\dv y_k[i-2]$. In block $i$, Encoder $k$ first looks for an appropriate compression index $t_{k,i-1}$ of $(\dv u_k[i-1],\dv y_k[i-1])$, and then sends $\dv x_{k,i}[i]:=\dv x_{k,i}(w_{k0},t_{k,i-1},w_{kk})$ generated on top of $\dv u_k[i]:= \dv u_{k,i}(w_{k0},t_{k,i-1})$. That is, the private message $w_{kk}$ is superimposed on top of the compression index $t_{k,i-1}$, which itself is superimposed on top of the public message $w_{k0}$. At the end of the transmission, Decoder $k$, $k=1,2$, has collected all its outputs $(\dv y_k[1],\dv y_k[2],\hdots,\dv y_k[B])$ from which it decodes jointly the pair  $(w_{k0},w_{kk})$ of public and private messages transmitted by the respective encoder, i.e., Encoder $k$, without uniquely decoding neither the compression indices from both encoders nor the public message from the other encoder.
\end{remark}

\vspace{-0.0cm}
\section{Injective Deterministic IC with Intermittent Feedback}\label{secIV}
\vspace{-0.1cm}

\begin{figure}[htpb]
\centering
\includegraphics[width=\linewidth]{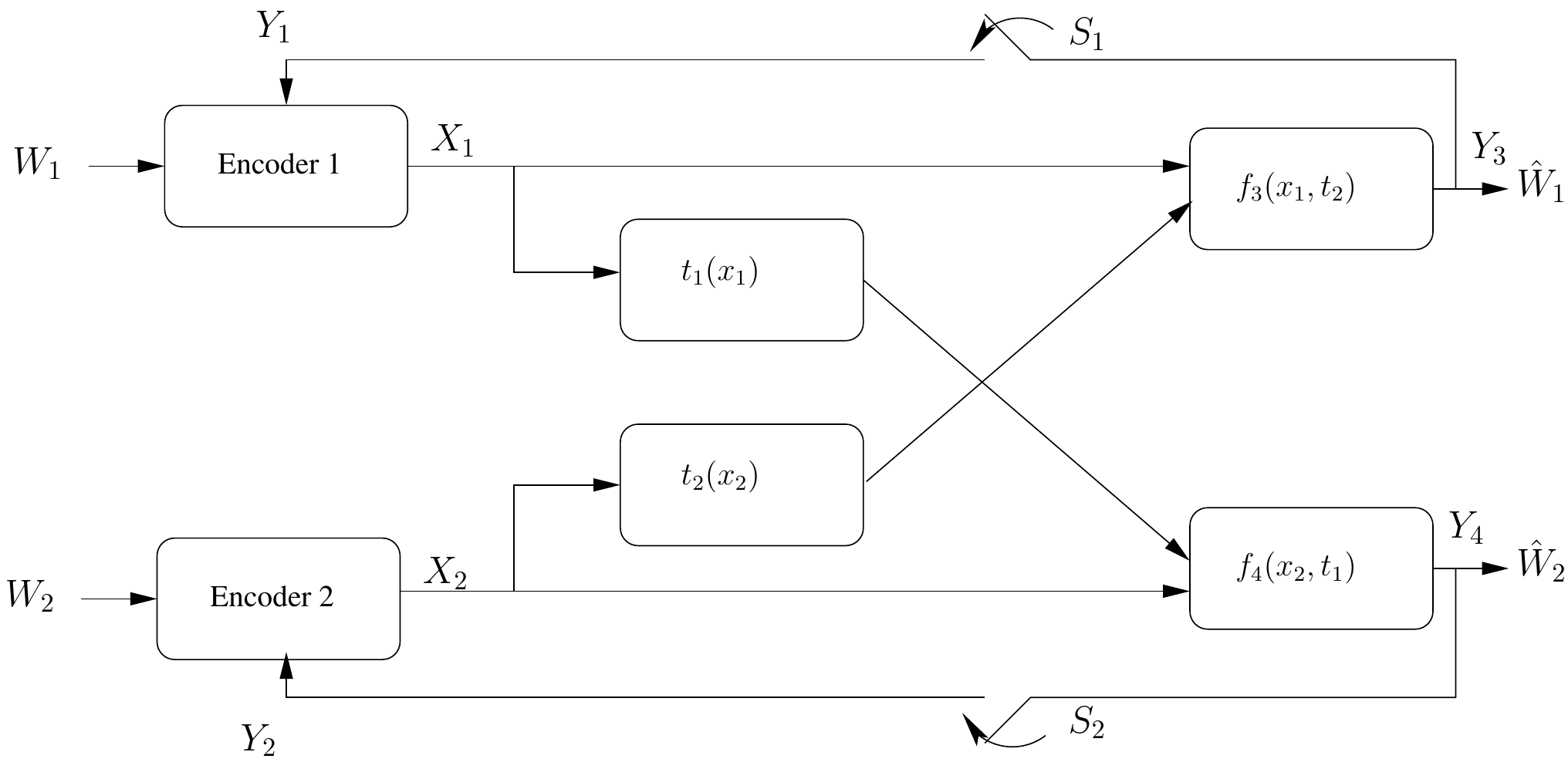}
\caption{Injective deterministic IC with intermittent feedback. Erasures in the feedback link from Decoder $k$ to encoder $k$, $k=1,2$, are governed by the state sequence $S^n_k$.}
\label{fig-injective-ic-if}
\end{figure}

Consider the deterministic interference channel depicted in Figure~\ref{fig-injective-ic-if}. The channel outputs are given by
\begin{equation}
Y_3 = f_3(X_1,T_2)\quad \text{and}\quad Y_4 = f_4(X_2,T_1)
\end{equation}
where $T_1=t_1(X_1)$ and $T_2=t_2(X_2)$ are functions of $X_1$ and $X_2$, respectively. We assume that the functions $f_3$ and $f_4$ are injective in $t_2$ and $t_1$, respectively. That is, for every $x_1 \in \mc X_1$, $f_3(x_1,t_2)$ is a one-to-one function of $t_2$, and for every $x_2 \in \mc X_2$, $f_4(x_2,t_1)$ is a one-to-one function of $t_1$. This class of interference channels is motivated by noiseless Gaussian IC, where the functions $f_3$ and $f_4$ are additions.

There are two feedback links, from Decoder $1$ to Encoder $1$ and from Decoder $2$ to Encoder $2$. We associate a state sequence $S^n_k$, $k=1,2$, with the feedback link from Decoder $k$ to Encoder $k$. The feedback state sequences are possibly correlated, with
\vspace{-0.2cm}
\begin{equation}
p_{S^n_1,S^n_2}(s^n_1,s^n_2) = \prod_{i=1}^n p_{S_1,S_2}(s_{1i},s_{2i}).
\end{equation}

Let $\{S_k[i]=1\}$, $k=1,2$, $i=1,\hdots,n$, denote the feedback event in the feedback link from Decoder $k$ to Encoder $k$ at time $i$, i.e., $Y_{k}[i]=Y_{k+2}[i]$ --- (the event $\{S_k[i]= *\}$ then denotes erasure in this feedback link at time $i$). We assume that 
\begin{equation}
\text{Pr}(S_1=1)=p_1\qquad \text{and}\qquad \text{Pr}(S_2=1)=p_2
\end{equation}
and that the state sequences are revealed in a strictly causal manner to \textit{both} encoders.

The input signal from Encoder $k$, $k=1,2$, at time $i$, $i=1,\hdots,n$, can depend on all the past observations at the encoder, i.e., $X_k[i] = \phi_{k,i}(W_k,S^{i-1}_1,S^{i-1}_2,Y^{i-1}_k)$. After $n$ channel uses, Decoder $k$, $k=1,2$, guesses an estimate $\hat{W}_k$ of message $W_k$ from its output $Y^n_{k+2}$.

\subsection{Inner Bound}\label{secIV_subsecA}

For convenience, we assume that  $*{\cdot}0=*{\cdot}1=*$. Let, for $k=1,2$, $\tilde{Y}_k=S_k{\cdot}Y_k$. At time $i$, we have $\tilde{Y}_k[i]=S_k[i]{\cdot}Y_k[i]$. Thus, $\tilde{Y}_k[i]={Y}_k[i]$ if $S_k[i]=1$ and $\tilde{Y}_k[i]=*$ if $S_k[i]=*$. With these notations, we have that, at time $i$, $i=2,\hdots,n$, Encoder $k$ gets $\tilde{Y}_k[i-1]$ by means of the intermittent feedback from Decoder $k$. Also, define $\tilde{T}_1=S_2{\cdot}T_1$ and $\tilde{T}_2=S_1{\cdot}T_2$. At time $i$, $\tilde{T}_1[i]=T_1[i]$ if $S_2[i]=1$ and  $\tilde{T}_1[i]=*$ if $S_2[i]=*$. Similarly, $\tilde{T}_2[i]=T_2[i]$ if $S_1[i]=1$ and  $\tilde{T}_2[i]=*$ if $S_1[i]=*$. Then, it is easy to see that because the function $f_3$ is injective in $t_2$ for every $x_1$, at time $i$, $i=1,\hdots,n$, Encoder $1$ also knows $\tilde{T}_2[i-1]$. Similarly,  because the function $f_4$ is injective in $t_1$ for every $x_2$, at time $i$ Encoder $2$ knows $\tilde{T}_1[i-1]$. Thus, at time $i$ each encoder knows both $\tilde{T}_1[i-1]$ and $\tilde{T}_2[i-1]$. Let $\tilde{T}=(\tilde{T}_1,\tilde{T}_2)=(S_2T_1,S_1T_2)$. The above means that, at time $i$, both encoders know the value of the pair $\tilde{T}[i-1]=(\tilde{T}_1[i-1], \tilde{T}_2[i-1])$.

Applying the coding scheme of Theorem~1 with the choice $Y_1=\tilde{T}_2$, $Y_2=\tilde{T}_1$, and $V_1=V_2=\tilde{T}=(\tilde{T}_1,\tilde{T}_2)$ we get the achievable region that will follow for the injective deterministic IC with intermittent feedback.

\vspace{0.3cm}

Define $\mc R^{\text{in}}_{\text{ic-if}}$ to be the set of all rate pairs $(R_1,R_2)$ such that there exist non-negative numbers $R_{10}$, $R_{20}$ satisfying
\begin{align}
R_1 &\leq H(Y_3|T_2,\tilde{T}_1,\tilde{T}_2,Q)+H(\tilde{T}_1|Q)\nonumber\\
R_1+R_{20} &\leq H(Y_3|Q)\nonumber\\
R_1-R_{10}+R_{20} &\leq \min\big\{H(Y_3|Q), H(Y_3|T_1,\tilde{T}_1,\tilde{T}_2,Q) + H(\tilde{T}_2|Q)\big\}\nonumber\\
R_1-R_{10}  &\leq \min\big\{H(Y_3|T_2,\tilde{T}_1,\tilde{T}_2,Q), H(Y_3|T_1,\tilde{T}_1,\tilde{T}_2,Q)\big\}\nonumber\\
R_1-R_{10} &\leq H(Y_3|T_1,T_2,\tilde{T}_1,\tilde{T}_2,Q)+H(\tilde{T}_1|Q)+H(\tilde{T}_2|Q)\nonumber\\
R_2 &\leq H(Y_4|T_1,\tilde{T}_1,\tilde{T}_2,Q)+H(\tilde{T}_2|Q)\nonumber\\
R_2+R_{10} &\leq H(Y_4|Q)\nonumber\\
R_2-R_{20}+R_{10} &\leq \min\big\{H(Y_4|Q), H(Y_4|T_2,\tilde{T}_1,\tilde{T}_2,Q) + H(\tilde{T}_1|Q)\big\}\nonumber\\
R_2-R_{20}  &\leq \min\big\{H(Y_4|T_1,\tilde{T}_1,\tilde{T}_2,Q), H(Y_4|T_2,\tilde{T}_1,\tilde{T}_2,Q)\big\} \nonumber\\
R_2-R_{20} &\leq H(Y_4|T_1,T_2,\tilde{T}_1,\tilde{T}_2,Q)+H(\tilde{T}_1|Q)+H(\tilde{T}_2|Q).
\label{inner-bound-ic-gf}
\end{align}

\vspace{-0.2cm}

\begin{theorem}\label{theorem-inner-bound-injective-ic-if}
The capacity region of the injective deterministic interference channel with intermittent feedback satisfies
\begin{equation}
\mc R^{\text{in}}_{\text{ic-if}} \subseteq \mc C_{\text{ic-if}}.
\end{equation}
\end{theorem}

\vspace{0.1cm}

\begin{remark}
Note that $T_1[i-1]$ is the interference that Transmitter $1$ causes to Receiver $2$ at time $i-1$. However, because the feedback from Decoder $2$ is intermittent, at time $i$ this interference is seen by Encoder $2$ as if it were (only) $\tilde{T}_1[i-1]$. Similarly, the interference $T_2[i-1]$ is seen by Encoder $1$ at time $i$ as if it were (only) $\tilde{T}_2[i-1]$. With the choice $V_1[i]=V_2[i]=(\tilde{T}_1[i-1],\tilde{T}_2[i-1])$, the coding scheme of Theorem~\ref{theorem-inner-bound-injective-ic-if} can be viewed as an \textit{appropriate}, non-trivial, combination of Lim \textit{et al.} noisy network coding \cite{H-LKGC11} or Avestimeher \textit{et al.} quantize-map-and-forward \cite{ADT11} and the technique of interference alignment.
\end{remark}

\subsection{Linear Deterministic IC with Intermittent Feedback}\label{secIV_subsecB}

In this section, we study an important special class of injective interference channels with intermittent feedback: the linear deterministic model \cite{ADT11} with intermittent feedback. Here, $\mc X_1=\mc X_2=\mathbb{F}^q_2$ . The input-output relations at time $i$, $i=1,\hdots,n$, are given by
\vspace{-0.2cm}
\begin{equation}
\dv Y_3[i] = \dv H_{11}\dv X_1[i] + \dv H_{12}\dv X_2[i], \quad \dv Y_4[i] = \dv H_{22}\dv X_2[i] + \dv H_{21}\dv X_1[i]
\label{input-output-relation-linear-deterministic-ic}
\vspace{-0.2cm}
\end{equation}
where $\dv X_k[i]$, $k=1,2$, are binary vectors of length $q$ denoting the transmitted signals at time (or block $i$); $\dv Y_{k+2}[i]$, $k=1,2$, are  binary vectors of same length denoting the received signals; channel matrices $(\dv H_{11},\dv H_{12},\dv H_{21},\dv H_{22})$ are such that, for  $(k,l) \in \{1,2\}^2$, $\dv H_{kl}=\dv H^{q-n_{kl}}$ with $\dv H$ denoting the $q{\times}q$ shift matrix and $n_{kl}$ are nonnegative numbers whose values are related to the channel gains. The summation in \eqref{input-output-relation-linear-deterministic-ic} is in $\mathbb{F}_2$ (modulo $2$). 



The inner bound of Theorem~\ref{theorem-inner-bound-injective-ic-if} also applies to the linear deterministic model \eqref{input-output-relation-linear-deterministic-ic} with intermittent feedback, a channel whose capacity region has been fully characterized recently in \cite[Theorem 3.1]{KWD13}. 

Define $\mc C$ be the set of all rate pairs $(R_1,R_2)$ satisfying
\begin{subequations}
\begin{align}
R_1 &\leq \min\:\{\max(n_{11},n_{12}),\: n_{11}+p_2(n_{21}-n_{11})^{+})\}\\
R_2 &\leq \min\:\{\max(n_{22},n_{21}),\: n_{22}+p_1(n_{12}-n_{22})^{+})\}\\
R_1+R_2 &\leq \min\:\big\{\max(n_{11},n_{12})+(n_{22}-n_{12})^{+},\nonumber\\
&\quad \max(n_{22},n_{21})+(n_{11}-n_{21})^{+}\big\}\\
R_1+R_2 &\leq \max\:\big\{n_{12},(n_{11}-n_{21})^{+}\big\} + \max\:\big\{n_{21},(n_{22}-n_{12})^{+}\big\}\nonumber\\
& + p_1\min\:\big\{n_{12},(n_{11}-n_{21})^{+}\big\} + p_2\min\:\big\{n_{21},(n_{22}-n_{12})^{+}\big\}\\
2R_1+R_2 &\leq \max(n_{11},n_{12}) + \max\:\big\{n_{21},(n_{22}-n_{12})^{+}\big\} \nonumber\\
& + (n_{11}-n_{21})^{+} + p_2\min\:\big\{n_{21},(n_{22}-n_{12})^{+}\big\}\\
R_1+2R_2 &\leq \max(n_{22},n_{21}) + \max\:\big\{n_{12},(n_{11}-n_{21})^{+}\big\} \nonumber\\
& + (n_{22}-n_{12})^{+} + p_1\min\:\big\{n_{12},(n_{11}-n_{21})^{+}\big\}.
\end{align}
\end{subequations}

\vspace{0.3cm}
As stated \cite[Theorem 3.1]{KWD13}, the set $\mc C$ characterizes the capacity region of the linear deterministic interference channel with intermittent feedback \eqref{input-output-relation-linear-deterministic-ic}.

\vspace{0.3cm}



\begin{remark}
The coding scheme of Theorem~\ref{theorem-inner-bound-injective-ic-if} (which, itself, can be obtained as an instance of that of Theorem~\ref{theorem-inner-bound-ic-gf}) is more general than that of \cite{KWD13}. At high level, however, the two coding schemes share elements, in that the interferences are aligned, compressed and then conveyed to the receivers. However, there are also substantial differences amon the two. In particular, by opposition to \cite{KWD13}, the compression indices are not uniquely decoded here; and simultaneous decoding is employed instead of backward decoding.
\end{remark}

\section{Proof of Achievability (Theorem~1)}\label{secV}

For convenience, we consider the case $Q=\emptyset$. Achievability for an arbitrary time-sharing random variable $Q$ can be proved using the coded time-sharing technique \cite{GK11}.

As we mentioned previously, the transmission takes place in $B$ blocks. Also, message $W_k$, $k=1,2$, is divided into a ``public" message $W_{k0}$ that is sent at rate $R_{k0}$ and a ``private" message $W_{kk}$ that is sent at rate $R_{kk}$. The total rate for message $W_k$ is then $R_k=R_{k0}+R_{kk}$.  The messages $(W_{10},W_{11},W_{20},W_{22})$ are sent over \textit{all} blocks. We thus have $B_{W_{k0}}=nB{R_{k0}}$, $B_{W_{kk}}=nB{R_{kk}}$, $N=nB$, $R_{W_{k0}}=B_{W_{k0}}/N=R_{k0}$ and $R_{W_{kk}}=B_{W_{kk}}/N=R_{kk}$, where $B_{W_{k0}}$ is the number of public message $W_{k0}$ bits, $B_{W_{kk}}$ is the number of private message $W_{kk}$ bits, $N$ is the total number of channel uses (over $B$ blocks) and $R_{W_{k0}}$ and $R_{W_{kk}}$ are the overall rates of the public message $W_{k0}$ and private message $W_{kk}$, respectively.

\noindent \textbf{Codebook Generation:} Fix a measure $P_{U_1,V_1,U_2,V_2,X_1,X_2,Y_1,Y_2,Y_3,Y_4} \in \mc P^{\text{in}}_{\text{ic-gf}}$. Fix $\epsilon > 0$, $\eta_{10} > 0$, $\eta_{11} > 0$, $\eta_{20} > 0$, $\eta_{22} > 0$, $\delta_1 > 0$, $\delta_2 > 0$ and denote $M_{10} = 2^{nB[R_{10}-\eta_{10}\epsilon]}$, $M_{11} = 2^{nB[R_{11}-\eta_{11}\epsilon]}$, $M_{20} = 2^{nB[R_{20}-\eta_{20}\epsilon]}$, $M_{22} = 2^{nB[R_{22}-\eta_{22}\epsilon]}$, $\hat{M}_1 = 2^{n[\hat{R}_1+\delta_1\epsilon]}$ and $\hat{M}_2 = 2^{n[\hat{R}_2+\delta_2\epsilon]}$.

\noindent We randomly and independently generate a codebook for each block.

\begin{itemize}
\item[1)] For each block $i$, $i=1,\hdots,B$, we generate $M_{10}\hat{M}_{1}$ independent and identically distributed (i.i.d.) codewords $\dv u_{1,i}(w_{10},t'_{1,i})$ indexed by $w_{10}=1,\hdots,M_{10}$, $t'_{1,i}=1,\hdots,\hat{M}_1$, each with i.i.d. components drawn according to $P_{U_1}$.

\noindent Similarly, for each block $i$, $i=1,\hdots,B$, we generate $M_{20}\hat{M}_{2}$ independent and identically distributed (i.i.d.) codewords $\dv u_{2,i}(w_{20},t'_{2,i})$ indexed by $w_{20}=1,\hdots,M_{20}$, $t'_{2,i}=1,\hdots,\hat{M}_2$, each with i.i.d. components drawn according to $P_{U_2}$.
\item[2)] For each block $i$, for each codeword $\dv u_{1,i}(w_{10},t'_{1,i})$,  we generate $\hat{M}_1$ i.i.d. codewords $\dv v_{1,i}(w_{10},t'_{1,i},t_{1,i})$ indexed by $t_{1,i}=1,\hdots,\hat{M}_1$, each with i.i.d. components drawn according to $P_{V_1|U_1}$.

\noindent Similarly, for each block $i$, for each codeword $\dv u_{2,i}(w_{20},t'_{2,i})$,  we generate $\hat{M}_2$ i.i.d. codewords $\dv v_{2,i}(w_{20},t'_{2,i},t_{2,i})$ indexed by $t_{2,i}=1,\hdots,\hat{M}_2$, each with i.i.d. components drawn according to $P_{V_2|U_2}$.
\item[3)]  For each block $i$, for each codeword $\dv u_{1,i}(w_{10},t'_{1,i})$, we generate $M_1$ i.i.d. codewords $\dv x_{1,i}=(w_{10},t'_{1,i},w_{11})\}$ indexed by $w_{11}=1,\hdots,M_1$, each with i.i.d. components draw according to $P_{X_1|U_1}$.

\noindent Similarly, for each block $i$, for each codeword $\dv u_{2,i}(w_{20},t'_{2,i})$, we generate $M_2$ i.i.d. codewords $\dv x_{2,i}=(w_{20},t'_{2,i},w_{22})\}$ indexed by $w_{22}=1,\hdots,M_2$, each with i.i.d. components draw according to $P_{X_2|U_2}$.
\end{itemize}

\textbf{Encoding:} Suppose that a message pair $(W_1,W_2)$ are to be transmitted, with $W_1=(w_{10},w_{11})$ and $W_2=(w_{20},w_{22})$. As we mentioned previously, $(w_{10},w_{11})$ and $(w_{20},w_{22})$ will be sent over \textit{all} blocks. We denote by $\dv y_k[i-1]$, $k=1,2$, $i=1,\hdots,B$, the GF output observed by Encoder $k$ in block $i$. For convenience, we let $\dv y_1[0]=\emptyset$ and $t_{1,-1}=t_{1,0}=1$ (a default value). Similarly, we let $\dv y_2[0]=\emptyset$ and $t_{2,-1}=t_{2,0}=1$. The encoding at the beginning of block $i$, $i=1,\hdots,B$, is as follows.

\noindent Encoder $1$ which has observed the GF output $\dv y_1[i-1]$, knows $t_{1,i-2}$ and looks for a compression index $t_{1,i-1} \in [1:\hat{M}_1]$ such that $\dv v_{1,i-1}(w_{10},t_{1,i-2},t_{1,i-1})$ is strongly jointly typical with $\dv y_1[i-1]$ and $\dv u_{1,i-1}(w_{10},t_{1,i-2})$. If there is no such index or the GF output $\dv y_1[i-1]$ is not typical, $t_{1,i-1}$ is set to $1$ and an error is declared. If there is more than one such index $t_{1,i-1}$, choose the smallest. One can show that the encoding error in this step is small as long as $n$ large and
\vspace{-0.2cm}
\begin{equation}
\hat{R}_1 > I(V_1;Y_1|U_1).
\label{constraint-encoding-at-encoder1}
\end{equation}
\noindent Encoder 1 transmits the vector $\dv x_{1,i}(w_{10},t_{1,i-1},w_{11})$. Similarly, Encoder 2 finds the appropriate compression index $t_{2,i-1}$ as long as $n$ large and
\vspace{-0.2cm}
\begin{equation}
\hat{R}_2 > I(V_2;Y_2|U_2).
\label{constraint-encoding-at-encoder2}
\end{equation}
Encoder $2$ then transmits the vector $\dv x_{2,i}(w_{20},t_{2,i-1},w_{22})$.

\textbf{Decoding:} We use simultaneous nonunique decoding. At the end of the transmission, Decoder $1$ has collected all the blocks of channel outputs $(\dv y_3[1],\hdots,\dv y_3[B])$ and Decoder $2$ has collected all the blocks of channel outputs $(\dv y_4[1],\hdots,\dv y_4[B])$.

\noindent Decoder $1$ estimates the pair $(w_{10},w_{11})$ using all blocks $i=1,\hdots,B$, i.e., simultaneous joint decoding. It finds the unique $(\hat{w}_{10},\hat{w}_{11})$ s.t. $\dv u_{1,i}(\hat{w}_{10},t_{1,i-1})$, $\dv v_{1,i}(\hat{w}_{10},t_{1,i-1},t_{1,i})$, $\dv x_{1,i}(\hat{w}_{10},t_{1,i-1},w_{1,1})$, $\dv u_{2,i}(w_{20},t_{2,i-1})$, $\dv v_{2,i}(w_{20},t_{2,i-1},t_{2,i})$, $\dv y_3[i]$ are jointly typical for all $i=1,\hdots,B$, for some $w_{20} \in [1,M_{20}]$ some compression indices $\dv t_1=(t_{1,1},\hdots,t_{1,B}) \in [1,\hat{M}_1]^B$ and $\dv t_2=(t_{2,1},\hdots,t_{2,B}) \in [1,\hat{M}_2]^B$; otherwise it declares an error. One can show that Decoder $1$ obtains the correct $(w_{10},w_{11})$ as long as $n$ and $B$ are large and
\begin{subequations}
\begin{align}
R_{11}+R_{10}+\hat{R}_1 &\leq I(U_1,V_1,X_1;Y_3,U_2,V_2)\\
R_{11}+R_{10}+\hat{R}_1+\hat{R}_2 &\leq I(U_1,V_1,U_2,V_2,X_1;Y_3) \nonumber\\
& + I(U_2,V_2;U_1,V_1,X_1)\\
R_{11}+R_{10}+R_{20}+\hat{R}_1+\hat{R}_2 &\leq I(U_1,V_1,U_2,V_2,X_1;Y_3) \nonumber\\
& + I(U_2,V_2;U_1,V_1,X_1)
\end{align}
\label{constraint-decoding-at-decoder1-step1}
\end{subequations}
and
\vspace{-0.3cm}
\begin{subequations}
\begin{align}
R_{11} &\leq I(X_1;Y_3|U_1,V_1,U_2,V_2) \nonumber\\
&+ I(U_2,V_2;U_1,V_1,X_1)\\
R_{11}+R_{20}+\hat{R}_2 &\leq I(U_2,V_2,X_1;Y_3|U_1,V_1) \nonumber\\
&+ I(U_2,V_2;U_1,V_1,X_1)\\
R_{11}+\hat{R}_1+\hat{R}_2 &\leq I(U_1,V_1,X_1;Y_3|U_2,V_2) \nonumber\\
& + I(U_2,V_2;U_1,V_1,X_1)\\
R_{11}+\hat{R}_1+\hat{R}_2 &\leq I(U_2,V_2,X_1;Y_3|U_1,V_1) \nonumber\\
& + I(U_2,V_2;U_1,V_1,X_1)\\
R_{11}+R_{20}+\hat{R}_1+\hat{R}_2 &\leq I(U_1,V_1,U_2,V_2,X_1;Y_3) \nonumber\\
&+ I(U_2,V_2;U_1,V_1,X_1).
\end{align}
\label{constraint-decoding-at-decoder1-step2}
\end{subequations}
Similarly, Decoder $2$ finds the correct pair $(w_{20},w_{22})$ if similar constraints, that are obtained by swapping the indices $1$ and $2$ and substituting the index $3$ with $4$ in \eqref{constraint-decoding-at-decoder1-step1} and \eqref{constraint-decoding-at-decoder1-step2}, hold.

\vspace{-0.1cm}

\bibliographystyle{IEEEtran}
\bibliography{ic-gf}
\end{document}